\documentclass[12pt]{article}

\usepackage[english]{babel}
\usepackage[T1]{fontenc}
\usepackage[utf8]{inputenc}
\usepackage{dsfont}

\usepackage{lmodern,microtype}

\usepackage{amsmath,amsthm,amssymb,amsfonts}

\usepackage{titlesec,titling}
\usepackage[nohead]{geometry}
\usepackage{setspace,caption}
\usepackage{enumitem,booktabs}
\usepackage{tabularx}
\usepackage{float}

\usepackage{graphicx}

\usepackage[comma,authoryear]{natbib}

\usepackage[dvipsnames]{xcolor}
\usepackage{hyperref}
\hypersetup{colorlinks=true,pdfnewwindow=true,pdfstartview=FitH,%
    pdftitle={Predicting Consumption Bundles with Chronos-2},%
    pdfauthor={Victor Aguiar and Nail Kashaev},%
    urlcolor=blue!90!red!45!black,citecolor=blue!90!red!45!black,linkcolor=red!90!black}

\usepackage{algorithm}
\usepackage{algpseudocode}

\newtheorem{definition}{Definition}

\DeclareCaptionFont{fancy}{\bfseries\sffamily}
\titleformat{\section}[block]{\centering\large\bfseries\sffamily}{\thesection.}{0.5em}{}
\titleformat{\subsection}[block]{\flushleft\bfseries\sffamily}{\thesubsection.}{0.5em}{}
\titleformat{\subsubsection}[runin]{\normalsize\itshape}{\bfseries\upshape\sffamily\thesubsubsection.}{0.5em}{}[.--\:]
\titlespacing{\section}{0ex}{10ex}{5ex}
\titlespacing{\subsection}{0in}{6ex}{3ex}

\allowdisplaybreaks

\providecommand{\Real}{{\mathds{R}}}

\title{GARP-EFM: Improving Foundation Models with Revealed Preference Structure}
\author{
Victor H.\ Aguiar\thanks{The views expressed in this paper are those of the authors and do not necessarily reflect the views of Amazon or any of its affiliates. This paper is provided for informational and research purposes only. Any errors are the authors' own.}\\
Amazon\\\texttt{vhaguiar@amazon.com}
\and
Nail Kashaev\\Western University\\\texttt{nkshaev@uwo.ca}
}
\date{\today}

\begin{document}
\maketitle

\begin{abstract}
Modern pretrained time-series foundation models can forecast without task-specific training, but they do not fully incorporate economic behavior. We show that teaching them basic economic logic improves how they predict demand using an experimental panel. We fine-tune Amazon Chronos-2, a transformer-based probabilistic time-series model, on synthetic data generated from utility-maximizing agents. We exploit Afriat's theorem, which guarantees that demand satisfies the Generalized Axiom of Revealed Preference (GARP) if and only if it can be generated by maximizing some utility function subject to a budget constraint. GARP is a simple condition to check that allows us to generate time series from a large class of utilities efficiently. The fine-tuned model serves as a rationality-constrained forecasting prior: it learns price-quantity relations from GARP-consistent synthetic histories and then uses those relations to predict the choices of real consumers. We find that
fine-tuning on GARP-consistent synthetic data substantially improves prediction relative to zero-shot Chronos-2 at all forecast horizons we study. Our results show that economic theory can be used to generate structured synthetic data that improves foundation-model predictions when the theory implies observable patterns in the data. 
\end{abstract}
\pagebreak

\section{Introduction}\label{sec: intro}
We study whether adding economic structure improves out-of-sample demand forecasting in pretrained time-series foundation models. These models can generate accurate predictions without task-specific training, but they rely purely on statistical patterns and may not explicitly encode how consumers respond to prices. Our approach is to train them on synthetic data generated from economic models, so that they learn economically meaningful relationships between prices and quantities.

We consider a setting in which a consumer faces a sequence of budget scenarios with exogenously varying prices and chooses bundles of multiple goods over time. The goal is to use past choices to predict future demand, making this a multivariate time-series problem over consumption bundles. Using Amazon Chronos-2, we first show that zero-shot forecasts outperform simple benchmarks. We then show that training on synthetic data generated by utility-maximizing agents leads to substantial additional improvements. Importantly, these gains arise not from fitting the observed sample, but from introducing a rationality-based prior that improves generalization in environments with rich price variation.

Two literatures have approached this problem from opposite directions. The structural approach assumes a particular specification of the utility function, estimates its parameters from observed choices, and derives predictions from the demand system implied by estimated utilities. This approach is interpretable and grounded in theory, but requires the data to be (approximately) consistent with the specification. In contrast, The machine learning approach imposes no structural restrictions. A foundation model trained on large time-series corpora produces accurate predictions from pattern recognition alone, without any model of the underlying agent.

Our approach combines the above two approaches. We fine-tune Amazon Chronos-2 on large synthetic panels generated from utility-maximizing agents. The synthetic data are never mixed with the real experimental observations. Instead, they serve as a structured prior that shapes the model's internal representations before it encounters the real data at test time. We use results from revealed preference literature \citep{varian1982nonparametric}, to do this efficiently and scalably using the Generalized Axiom of Revealed Preference (GARP). GARP is a condition that can be easily checked. Moreover, it allows us to easily generate data from any utility-maximizing behavior. 

The empirical panel we use to test the models comes from the portfolio-choice experiment of \citet{ahn2014estimating}. It contains choices from $N = 154$ consumers across $T = 50$ budget scenarios. We choose it because it is a non-trivial forecasting task with very high-experimental price variation, long time-series and 3 prices and 3 goods. The panel is far from being perfectly rationalized by utility-maximizing behavior: only
20 of 154 consumers satisfy GARP on their full 50-period choice history. This is precisely the setting in which a synthetic prior may be useful. We do not claim that consumers are literally GARP-consistent agents. In fact, the overwhelmingly majority of them are not consistent with GARP. Rather, we ask whether training on histories that are coherent with utility maximizing behavior helps predict the choices of agents who are only approximately rational. \citet{varian1982nonparametric} showed that when consumers are consistent with GARP we can use GARP alone to produce set-valued out-of-sample predictions for demand given prices. That approach is highly efficient for the few cases where consumers are exactly rational. When consumers are not rational there is a literature on working on interpolation with regularization using the restrictions imposed on the demand function by rationality \citep{aguiar2018classifying,aguiar2017slutsky}. Recently, and in a similar vein, \citet{andrewsrevealed} proposes a regularization based on GARP and measures of goodness-of-fit for transformer-based architectures. Albeit attractive theoretically, the proposal in \citet{andrewsrevealed} has several potential drawbacks. First, the proposed regularization may affect the properties of optimizers in unexpected ways. Second, the regularization using GARP requires the choice of tuning parameter. Finally, recent evidence questions the reliability of goodness-of-fit rationality measures. \citet{Nitsch2022ReliabilityRationality} shows that widely used indices such as the Critical Cost Efficiency Index (CCEI), proposed in \citet{andrewsrevealed} as a target for regularization, are highly sensitive to measurement design, limiting their usefulness as stable characterizations of decision-making behavior. In contrast, we depart from the measurement paradigm altogether. Rather than estimating or testing rationality at the individual level, we use revealed-preference theory to generate synthetic data and fine-tune a foundation model on economically coherent behavior. This shifts the role of revealed preference from a diagnostic tool to a generative training principle, embedding rational structure directly into the model's representations. 

We show that fine-tuning on GARP-consistent synthetic data reduces bundle prediction error by 17-18\% relative to zero-shot Chronos-2 at forecast horizons of $H=5$, $10$, and $15$ and by 31\%  at $H=1$. 

The paper makes two contributions. The first is methodological: we provide a concrete channel through which economic theory can improve foundation-model forecasting without structural estimation. GARP-consistent synthetic data simulation is highly computational efficient and scalable for different number of time periods and very large number of goods. The second is substantive: we show that revealed-preference consistency is rich enough to generate a useful synthetic training distribution for forecasting real demand paths.

The remainder of the paper is organized as follows. Section~\ref{sec: cronos} describes Chronos-2 and how it generates forecasts. Section~\ref{sec: finetuning} explains LoRA fine-tuning. Section~\ref{sec: theory} defines GARP and details the synthetic data generation process. Section~\ref{sec: data} describes the experimental panel. Section~\ref{sec: setup} states the synthetic-panel parameters and fine-tuning configuration. Section~\ref{sec: results} reports the results and discusses the findings. We conclude in Section~\ref{sec: conclusion}.

\section{Chronos-2 for predicting customer behavior}\label{sec: cronos}

Chronos-2 (\texttt{amazon/chronos-2}) is a transformer-based probabilistic forecasting model. Its core architecture uses a T5-style, encoder-only configuration. Given a context window of length $L$, the \texttt{amazon/chronos-2} checkpoint we use outputs a predictive distribution. In multivariate mode, the model accepts a dictionary input with three fields: (i) \texttt{target}: a $(K \times L)$ array of historical values for $K$ consumption goods (e.g., $x$, $y$, $z$) over $L$ periods;
 (ii) \texttt{past\_covariates}: a dictionary mapping covariate names to $L$ arrays observed over the context window (past prices);
 (iii) \texttt{future\_covariates}: a dictionary mapping covariate names to $H$ arrays known over the forecast horizon (future prices). Goods are forecast \emph{jointly}: the prediction for good $x$ at period $t$ is based on the full history of goods $y$ and $z$ and all price series. Prices are particularly important covariates in this application because subjects observe future prices before making their choices, so future prices are known at forecast time.

Chronos-2 does not forecast by evaluating a closed-form demand function. Instead, it processes the input through four sequential transformations.
\paragraph{Instance normalization.}
Each series is normalized by its own context mean and standard deviation before being processed. This removes level differences between goods and subjects, allowing the model to focus on relative dynamics.
\paragraph{Temporal patching.}
The normalized series are divided into short overlapping temporal patches. Each patch captures a local segment of the time series and is associated with a time encoding that records its position relative to the forecast origin. Past patches receive negative time indices; future covariate patches receive nonnegative indices. This explicit temporal encoding allows the model to distinguish observations in the historical window from those in the forecast horizon.
\paragraph{Attention.}
The patch embeddings are processed by two kinds of attention. \emph{Time attention} links each patch to other temporal positions in the same forecasting problem, allowing the model to capture long-range dependencies in the context window. In multivariate mode, \emph{group attention} uses subject-level identifiers to construct an attention mask: the three goods of the same subject can exchange information, but goods belonging to different subjects in the same batch do not interact. This is the mechanism through which cross-good price and quantity patterns within a subject are jointly captured.

\paragraph{Output head.}
The final hidden states associated with the future horizon positions are passed through an output patch head that returns 21 quantile forecasts for each period in the horizon. The median quantile (level 0.50) is used as the point forecast in the MASE and bundle $\ell_2$ evaluations; the full quantile distribution is used for the Weighted Quantile Loss.

\subsection{Why Fine-Tuning on Synthetic Data Helps}

Chronos-2 is pre-trained on a large set of different time series. It captures generic temporal patterns --trends, seasonality, mean reversion-- but not the economic structure of consumer demand. In particular, the pre-trained model need not capture cross price elasticity. That is, the model has no reason to associate movements in the price of good $x$ with changes in demands for goods $y$ and $z$ under a fixed budget. 

Fine-tuning on synthetic data generated from utility-maximizing agents reshapes the model's representations so that price-responsive demand patterns are encoded in its embeddings and attention weights. Utility maximization imposes a coherent substitution structure across goods, linking their forecasts through the budget constraint. 

After fine-tuning, when the model encounters a real subject's context window, it can match that history to economically meaningful patterns in its training data and produce forecasts that reflect the qualitative structure of demand. Importantly, this mechanism does not require consumers to be exactly rational: synthetic rational histories act as a structured prior over price-quantity sequences, improving predictions even when observed behavior only approximately satisfies rationality.

\paragraph{Interpretability and Recovering Economic Structure.}

An important question is whether the economic model induced by GARP-consistent synthetic data can be recovered from the trained model. Our approach does not explicitly estimate a utility function, nor does it impose parametric restrictions on preferences. However, by construction, the model is trained on data generated by utility-maximizing agents, and therefore its predictions inherit qualitative properties of rational demand.

This suggests a form of implicit interpretability: the learned representations encode substitution patterns that are consistent with utility maximization. Rather than recovering a utility function directly, one can evaluate the extent to which the model's predictions satisfy revealed-preference conditions or approximate them in the sense of Afriat efficiency. In this way, revealed preference provides a diagnostic tool for interpreting the model's behavior.

This perspective complements existing approaches that impose economic models through constraints or penalties. \citet{aguiar2018classifying} characterize how to do demand interpolation regularizing the deviations from the Slutsky (substitution matrix) restrictions, while  \citet{andrewsrevealed} proposes incorporating revealed-preference conditions as a regularization term in the training objective of transformer models. In contrast, our approach introduces economic models through the training distribution itself: by exposing the model to GARP-consistent synthetic histories, we shape its representations ex ante rather than penalizing deviations ex post. 

These approaches are complementary. Regularization enforces rationality locally during training, while synthetic-data training provides a global prior over economically coherent behavior. Together, they suggest a broader framework in which economic theory can be used both to guide learning and to interpret the resulting models.

\section{Fine-Tuning with LoRA}\label{sec: finetuning}

\subsection{Low-Rank Adaptation}

Full fine-tuning of Chronos-2 (originally $120M$ parameters) on a dataset of $\sim 50000$ agents $\times$ $50$ periods would risk overfitting and is computationally costly. The pre-trained model already possesses strong sequence-modelling ability; we only need to adapt its representations to the price-quantity structure of consumer demand. We therefore use Low-Rank Adaptation \citep[LoRA,][]{hu2021lora}, which keeps the original weights frozen and inserts a pair of trainable low-rank matrices into each attention projection.

Formally, for a pre-trained weight matrix $W_0 \in \Real^{d \times k}$, LoRA parameterizes the adapted weight as
\[
 W = W_0 + BA, \qquad B \in \Real^{d \times r},\; A \in \Real^{r \times k},
\]
where the rank $r \ll \min\{d, k\}$. Only $B$ and $A$ are updated during fine-tuning. This reduces the number of trainable parameters to roughly 1-2\% of the total.

\paragraph{For economists.}
Think of LoRA as a restricted re-estimation strategy. The pre-trained Chronos-2 is a high-dimensional general-purpose prior. LoRA adds a small number of task-specific parameters that perturb a few key linear maps inside the network. The advantage is the same as in classical regularized estimation: lower variance, less overfitting, and a cheaper optimization problem.

\subsection{Training Configuration}

\begin{center}
\begin{tabular}{ll}
\toprule
Hyperparameter & Value \\
\midrule
Method      & LoRA \\
Steps       & 5{,}000 \\
Learning rate   & $10^{-5}$ (cosine schedule) \\
Batch size    & 64 series \\
Context length  & 35 periods \\
Prediction length & 15 periods \\
Optimiser     & AdamW \\
Precision     & \texttt{bfloat16} \\
Device      & Apple M2 Ultra (128 GB), MPS \\
\bottomrule
\end{tabular}
\end{center}

For each synthetic panel, we train on $40,000$ agents ($80$\% split) and validate on $10,000$. At each step, the data loader randomly samples a subsequence from each agent's $50$-period series, providing implicit data augmentation.

\section{Utility Maximization and Synthetic Data Generation}\label{sec: theory}

A consumer faces a sequence of budget problems $\Real$. At prices $p_t \in \Real^K_+$ and budget $m > 0$, the budget set is
\[
B(p_t, m) = \{q \in \Real^K_+ : p_t \cdot q \leq m\}.
\]
We observe a finite dataset of price-quantity pairs $\{p_t, q_t\}_{t=1}^T$, where each $q_t\in B(p_t, m)$. 

In the paper's main specification, the synthetic training data are generated from histories that satisfy GARP. This section states the rationality condition and the algorithm used to generate such synthetic panels.

We say that bundle $q_s$ is \emph{directly revealed preferred} to $q_t$ if $q_t$ is affordable when $q_s$ is chosen, that is, $p_s\cdot q_t \leq m$, written $q_s \succeq^{R,D} q_t$. We write $q_s \succ^{R,D} q_t$ if, in addition, $p_s\cdot q_t < m$. Let $\succeq^R$ denote the transitive closure of  $\succeq^{R,D}$: thus $q_s \succeq^R q_t$ if there is a chain $q_s \succeq^{R,D} q_{u_1} \succeq^{R,D} \cdots \succeq^{R,D} q_t$.

\begin{definition}[GARP]
A dataset $\{p_t,q_t\}_{t=1}^T$ satisfies the Generalized Axiom of Revealed Preference (GARP) if there is no $s$ and $t$ such that
 \[
 q_s \succeq^R q_t
 \quad\text{and}\quad
 q_t \succ^{R,D} q_s.
 \]
\end{definition}

The standard rationalization notion is utility maximization: a utility function $u:\Real_+^L\to\Real$ rationalizes the dataset if each observed bundle $q_t$ solves
 \[
 \max_{q\in B(p_t,m)} u(q).
 \]

Afriat's theorem shows that a dataset is consistent with GARP if and only if some locally non-satiated utility function rationalizes that dataset.\footnote{A utility function $u$ is \emph{locally non-satiated} if, at every bundle, there are arbitrarily small perturbations that are strictly preferred. Formally, for every $q\in\Real_+^L$ and every $\varepsilon>0$, there exists $q'\in\Real_+^L$ such that $\|q'-q\|<\varepsilon$ and $u(q')>u(q)$.} This means that if we generate a synthetic dataset that is consistent with GARP we are able to sample from a very large space of utility functions.

\paragraph{Data generation.}
We generate GARP-consistent synthetic panels using \texttt{simGarpPriceWealth} from \texttt{revealedPrefs}, a modified fork of \citet{boelaert2014revealedprefs} developed for \citet{aguiar2021stochastic}. For each period the sampler draws a candidate bundle and accepts it only if adding it to the sequence does not create a GARP violation.

\begin{algorithm}[H]
\caption{Full GARP data generation (per agent)}
\begin{algorithmic}[1]
\Require Prices $\{p_t\}_{t=1}^T$, budget $m$, max iterations $M$
\State $\mathcal{H} \leftarrow \emptyset$
\For{$t = 1, \ldots, T$}
 \Repeat
  \State Draw $(s_1, s_2, s_3) \sim \mathrm{Dirichlet}(1,1,1)$
  \State Set $\tilde{q}_{tk} = s_k \cdot m / p_{tk}$ for $k = 1,2,3$
  \Comment{$p_t \cdot \tilde{q}_t = m$ exactly}
 \Until{$\mathcal{H} \cup \{(p_t,\tilde{q}_t)\}$ satisfies GARP, or $M$ draws exhausted}
 \State $\mathcal{H} \leftarrow \mathcal{H} \cup \{(p_t, \tilde{q}_t)\}$
\EndFor
\State \Return $\mathcal{H}$
\end{algorithmic}
\end{algorithm}

GARP is verified at each step by a recursive depth-first search in C++ (RcppArmadillo). The Dirichlet proposal draws budget shares uniformly over the simplex, so the resulting panel covers a wide range of interior demand patterns. This sampler explores the full GARP polytope.

\section{Experimental Data}\label{sec: data}

We use the empirical panel from the portfolio-choice experiment of \citet{ahn2014estimating}. The dataset contains choices from $N = 154$ consumers across $T = 50$ budget scenarios. For each consumer $i$, we observe
\[
 D_i = \{p_{it}, q_{it}\}_{t=1}^{50},
 \qquad p_{it}, q_{it} \in \Real^3_+,
\]
where $p_{it} = (p^x_{it}, p^y_{it}, p^z_{it})$ are prices set by the experimenter and $q_{it} = (x_{it}, y_{it}, z_{it})$ is the chosen bundle. The budget is $m = 100$ in every scenario.

\paragraph{Exogeneity.}
Prices are set by the experimenter rather than determined in a market equilibrium. This removes the standard endogeneity concern that price movements are correlated with unobserved demand shocks. For our purposes, this makes the experiment an especially clean sandbox: future prices are genuinely known at prediction time, so the model can treat them as exogenous covariates exactly as revealed-preference analysis intends.

\paragraph{Rationality of the sample.}
Applying the GARP test to each consumer's full 50-period choice history, only 20 of 154 consumers (13\%) satisfy GARP. The mean CCEI is 0.870. The empirical panel is therefore far from being perfectly rationalized.

This motivates our approach. Rather than fitting a structural model to noisy individual choices, we use synthetic utility-maximizing panels as a training prior for the forecasting model. The fine-tuned model does not require real consumers to be exactly rational; it only needs the price-quantity patterns in rational histories to resemble those in real choice data.

\paragraph{Train-test split.}
For each consumer, the first 35 observations form the context window and the last 15 are held out. We report results for four forecast horizons $H \in \{1, 5, 10, 15\}$, each using the same 35-period context.

\paragraph{Revealed preference for ML readers.}
\citet{afriat1967construction} and \citet{varian1982nonparametric} show that GARP is equivalent to the existence of a utility function that rationalizes the full dataset. GARP is not a local shape restriction; it is a global coherence condition across all observed price-quantity pairs. If a consumer fails GARP, no utility-maximizing story is consistent with their entire choice history.

\section{Experimental Setup}\label{sec: setup}

\subsection{Synthetic Panel Parameters}

The synthetic GARP panel shares the same price DGP as the main application. Each price is drawn independently as
\[
 p^k_{it} \;\sim\; \mathrm{LogNormal}(\log 3,\; 0.5), \qquad k \in \{x, y, z\},
\]
giving prices roughly in $[1, 10]$ with mean $\approx 3.4$. The budget is
$m = 100$ in all panels, matching the experimental data. Each synthetic agent is observed over $T = 50$ periods.

\begin{center}
\begin{tabular}{llcc}
\toprule
Model & Total agents & Agents per price path & Price paths \\
\midrule
GARP 50K & 50{,}000 & 50{,}000 & iid per agent \\
\bottomrule
\end{tabular}
\end{center}

Each synthetic GARP agent draws its own independent price path.

\subsection{Fine-Tuning Hyperparameters}

The fine-tuned GARP model uses the LoRA configuration described in Section~\ref{sec: finetuning}. The training-validation split is 80/20 by agent. The context and prediction lengths during training (35 and 15 periods) match the evaluation protocol.

\begin{center}
\begin{tabular}{ll}
\toprule
Parameter & Value \\
\midrule
LoRA rank       & 16 \\
LoRA $\alpha$     & 32 \\
Steps         & 5{,}000 \\
Learning rate     & $10^{-5}$ \\
LR schedule      & cosine decay \\
Batch size      & 64 \\
Optimiser       & AdamW \\
Precision       & \texttt{bfloat16} \\
\bottomrule
\end{tabular}
\end{center}

\subsection{Evaluation Metrics}
We use several evaluation metrics to assess the performance of the models.
\paragraph{MASE.}
The Mean Absolute Scaled Error scales the forecast Mean Absolute Error (MAE) by the na\"ive random-walk MAE on the context:
\[
 \mathrm{MASE}_k=\frac{\frac{1}{H}\sum_{t=1}^{H}|q_{tk} - \hat{q}_{tk}|}{\frac{1}{L-1}\sum_{t=2}^{L}|q_{tk} - q_{t-1k}|}.
\]
Values below 1 indicate that the model outperforms the na\"ive baseline. We compute MASE for each good $k$.

\paragraph{Bundle $\ell_2$.}
The mean per-period Euclidean distance between the predicted and actual bundle:
\[
 \mathrm{L2} = \frac{1}{H}\sum_{t=1}^{H}\|\hat{q}_t - q_t\|.
\]
This is the primary metric for comparing joint bundle predictions.

\paragraph{Bundle fitness.}
For consumer $i$ and model $m$, the normalized bundle $\ell_2$ is
\[
 \mathrm{nL2}_i^m
 =
 \frac{\frac{1}{H}\sum_{t=1}^{H}\|\hat{q}_{it}^m - q_{it}\|_2}
    {\frac{1}{H}\sum_{t=1}^{H}\|q_{it}\|_2},
\]
and bundle fitness is $\mathrm{BF}_i^m = 1 - \mathrm{nL2}_i^m$. This metric equals 1 under perfect prediction and decreases as the normalized bundle miss grows. We report it alongside the CCEI scatter plot to link predictive accuracy with revealed-preference consistency.

\section{Results}\label{sec: results}

We report out-of-sample accuracy on the 154-consumer experimental panel for four forecast horizons. The context window is always 35 periods and the hold-out set is the next $H$ periods. We compare three specifications: Zero-shot (vanilla Chronos-2 with no fine-tuning) and GARP 50K (LoRA fine-tuned on the full synthetic GARP panel). We also include a random feasible-budget benchmark in the spirit of \citet{bronars1987power}: for each hold-out period, a bundle is drawn uniformly from the budget simplex by setting $\tilde{q}_k = s_k m / p_k$, where $(s_1, s_2, s_3) \sim \mathrm{Dirichlet}(1,1,1)$. This benchmark satisfies the budget constraint exactly but ignores all information in the context; MASE values above one confirm that it is outperformed by the na\"{i}ve random-walk baseline. 

In the Tables~\ref{tab: h=1}-\ref{tab: h=15}, we report MASE by good, average across consumers bundle $\ell_2$, and average across consumers bundle fitness for the three specifications. Fine-tuning on GARP-consistent synthetic data consistently improves over
zero-shot Chronos-2 at all four horizons. Bundle $\ell_2$ falls from
20.53 to 14.09 at $H = 1$, from 16.97 to 14.04 at $H = 5$, from 16.83 to
13.92 at $H = 10$, and from 17.29 to 14.23 at $H = 15$.

\begin{table}[ht]
\centering
\caption{MASE by good, bundle $\ell_2$, and bundle fitness, horizon $H = 1$ (lower is better for MASE and bundle $\ell_2$; higher is better for fitness).}
\begin{tabular}{lccccc}
\toprule
Model & $x$ & $y$ & $z$ & Bundle $\ell_2$ & Bundle fitness \\
\midrule
Random feasible budget & 1.200 & 1.171 & 1.362 & 31.48 & 0.274 \\
\addlinespace
Zero-shot       & 0.695 & 0.714 & 0.702 & 20.53 & 0.584 \\
\addlinespace
GARP 50K        & 0.498 & 0.449 & 0.464 & \textbf{14.09} & \textbf{0.683} \\
\bottomrule
\end{tabular}
\label{tab: h=1}
\end{table}

\begin{table}[ht]
\centering
\caption{MASE by good, bundle $\ell_2$, and bundle fitness, horizon $H = 5$ (lower is better for MASE and bundle $\ell_2$; higher is better for fitness).}
\begin{tabular}{lccccc}
\toprule
Model & $x$ & $y$ & $z$ & Bundle $\ell_2$ & Bundle fitness \\
\midrule
Random feasible budget & 1.206 & 1.130 & 1.268 & 29.98 & 0.292 \\
\addlinespace
Zero-shot       & 0.546 & 0.555 & 0.583 & 16.97 & 0.634 \\
\addlinespace
GARP 50K        & 0.464 & 0.477 & 0.511 & \textbf{14.04} & \textbf{0.690} \\
\bottomrule
\end{tabular}
\label{tab: h=5}
\end{table}

\begin{table}[ht]
\centering
\caption{MASE by good, bundle $\ell_2$, and bundle fitness, horizon $H = 10$ (lower is better for MASE and bundle $\ell_2$; higher is better for fitness).}
\begin{tabular}{lccccc}
\toprule
Model & $x$ & $y$ & $z$ & Bundle $\ell_2$ & Bundle fitness \\
\midrule
Random feasible budget & 1.234 & 1.127 & 1.231 & 29.75 & 0.292 \\
\addlinespace
Zero-shot       & 0.584 & 0.542 & 0.575 & 16.83 & 0.635 \\
\addlinespace
GARP 50K        & 0.498 & 0.458 & 0.495 & \textbf{13.92} & \textbf{0.692} \\
\bottomrule
\end{tabular}
\label{tab: h=10}
\end{table}

\begin{table}[ht]
\centering
\caption{MASE by good, bundle $\ell_2$, and bundle fitness, horizon $H = 15$ (lower is better for MASE and bundle $\ell_2$; higher is better for fitness).}
\begin{tabular}{lccccc}
\toprule
Model & $x$ & $y$ & $z$ & Bundle $\ell_2$ & Bundle fitness \\
\midrule
Random feasible budget & 1.233 & 1.141 & 1.242 & 29.95 & 0.295 \\
\addlinespace
Zero-shot       & 0.609 & 0.571 & 0.588 & 17.29 & 0.629 \\
\addlinespace
GARP 50K        & 0.520 & 0.478 & 0.512 & \textbf{14.23} & \textbf{0.690} \\
\bottomrule
\end{tabular}
\label{tab: h=15}
\end{table}

\begin{figure}[H]
\centering
\includegraphics[width=0.54\textwidth,height=0.34\textheight,keepaspectratio]{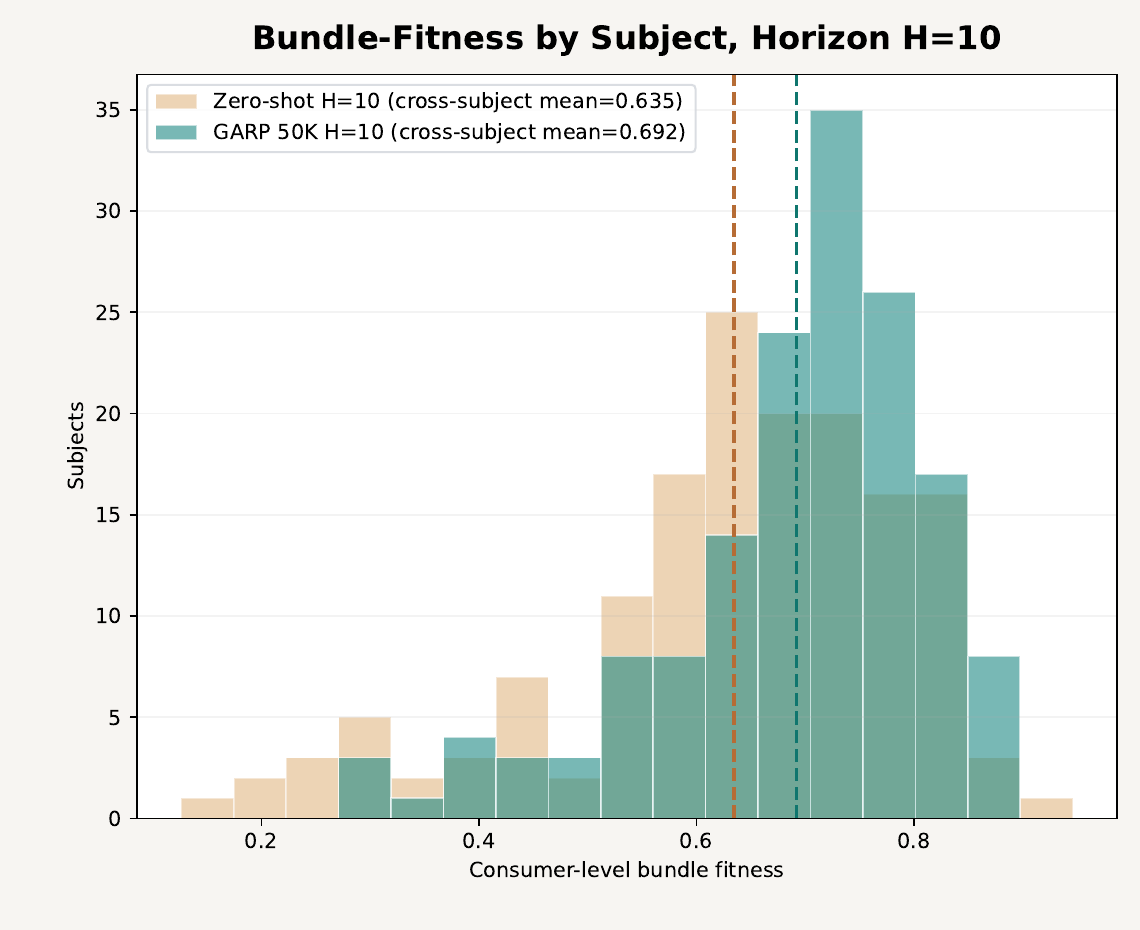}
\caption{Histogram of consumer-level bundle fitness at horizon $H = 10$,
   comparing zero-shot Chronos-2 with GARP 50K.
   Higher values are better.}
\label{fig: fitness-hist}
\end{figure}

Next we plot two figures to see the detailed comparison between the GARP-tuned model and the zero-shot Chronos-2 at horizon $H=10$. Figure~\ref{fig: fitness-hist} depicts a histogram of bundle fitness across consumers. At horizon $H = 10$, mean bundle fitness rises from 0.635 under zero-shot to 0.692 for GARP 50K, while the median rises from 0.656 to 0.717. The GARP model improves on zero-shot for 121 of the 154 consumers. Exact GARP passers ($20/154\approx 0.13$) have mean bundle fitness $0.723$ versus $0.688$ for the remaining consumers. For 33 of the 154 ($\approx 21\%$) consumers, the GARP model has lower bundle fitness than zero-shot.

\begin{figure}[H]
\centering
\includegraphics[width=0.84\textwidth,height=0.34\textheight,keepaspectratio]{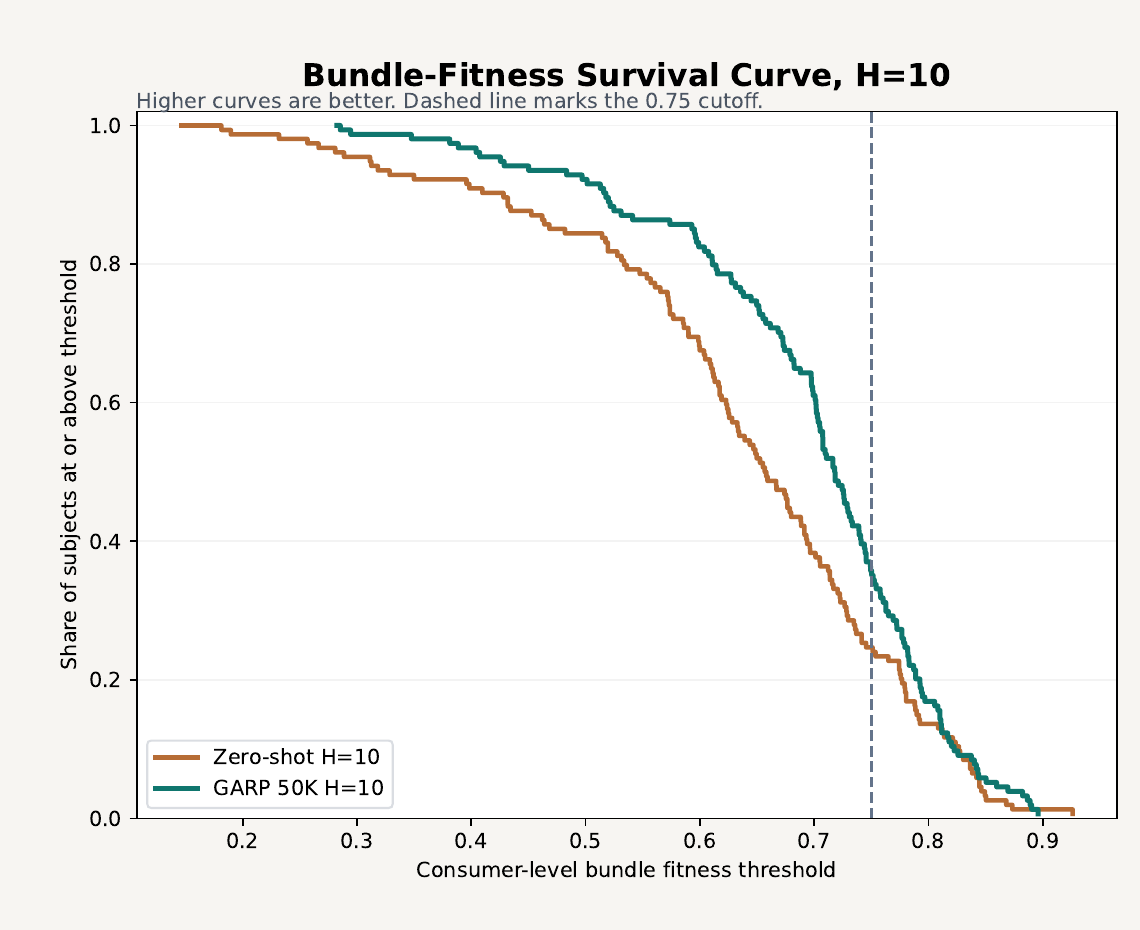}
\caption{Share of consumers achieving at least each bundle-fitness threshold at
   $H = 10$. The dashed line marks the 0.75 cutoff.}
\label{fig: fitness-threshold}
\end{figure}

In Figure~\ref{fig: fitness-threshold}, we plot the survival curves --the share of consumers whose bundle fitness is at or above the given level-- for both models. It shows that the improvement is broad-based. At $H = 10$, at almost every threshold the survival curve of the fine-tuned model dominates the zero-shot one. At the 0.75 cutoff, the share of consumers above threshold rises from $37/154\approx 0.24$ under zero-shot to $54/154\approx 0.35$ under the GARP-tuned model.

This pattern is consistent with the view that synthetic rational histories provide a useful prior over price-quantity relations. The pre-trained model processes many time series but few, if any, exhibiting the budget-constrained substitution patterns present in demand data. Fine-tuning on synthetic demand panels updates the model's internal representations in the direction of these patterns, and this update carries over to real consumers even when they are not perfectly rational.

\begin{figure}[H]
\centering
\includegraphics[width=0.84\textwidth,height=0.36\textheight,keepaspectratio]{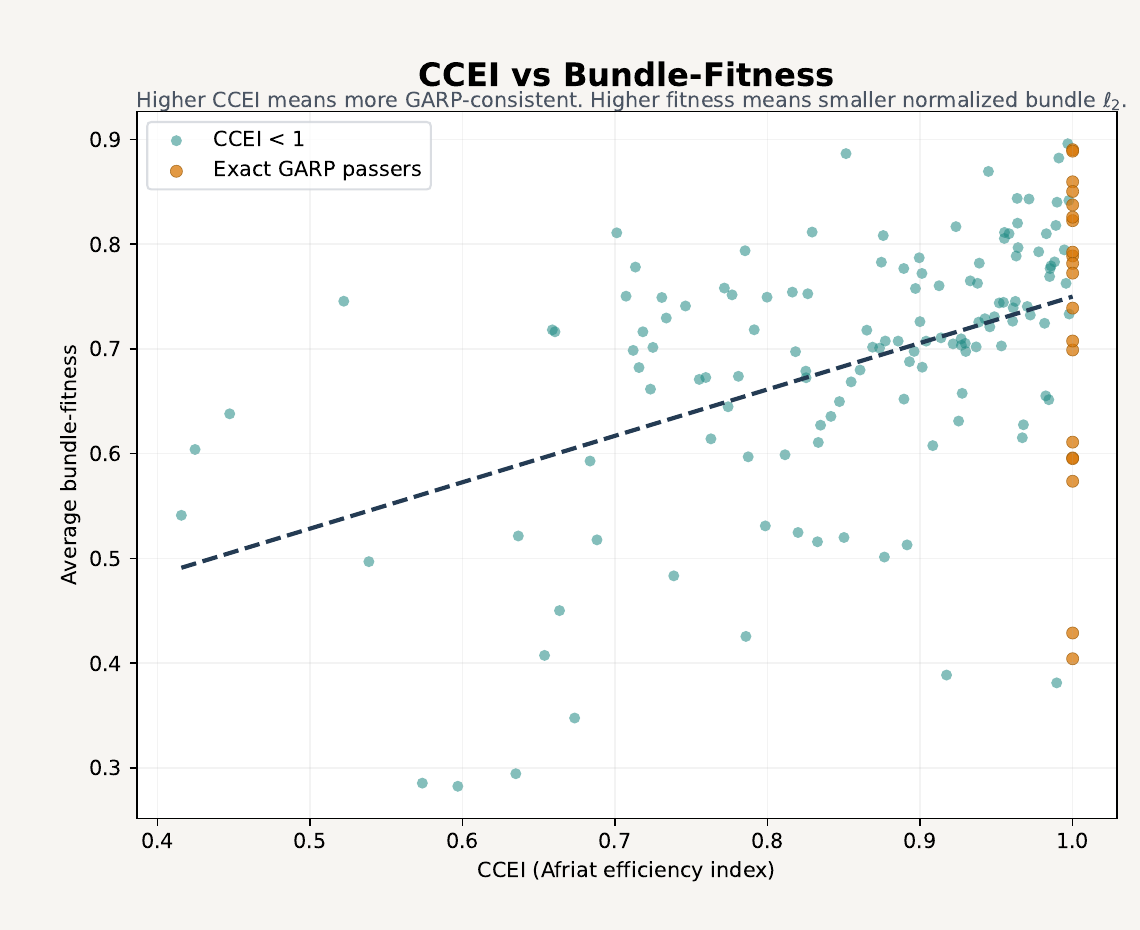}
\caption{Consumer-level CCEI against bundle fitness for GARP 50K at
   $H = 10$. Orange markers are exact GARP passers (CCEI $= 1$).}
\label{fig: ccei-fitness}
\end{figure}

Finally, we investigate the relationship between the bundle fitness of the fine-tuned model and the level of rationality measured by CCEI. In Figure~\ref{fig: ccei-fitness}, we scatterplot consumer bundle fitness against their CCEI at horizon $H=10$. 
The correlation between CCEI and bundle fitness is 0.456. Exact GARP passers (20/154) have mean bundle fitness 0.723 versus 0.688 for the remaining consumers, confirming that on average the choices of more rational consumers are predicted better by the GARP-tuned model. However, the correlation is not high enough to make CCEI a reliable screening rule (the GARP-tuned model underperforms zero-shot for 33 of the 154 consumers). The simpler strategy --applying the GARP model to all consumers-- is likely preferable in practice.

\section{Conclusion}\label{sec: conclusion}

We show that utility maximizing behavior can be incorporated into a probabilistic time-series foundation model without structural estimation on the target population or direct regularization. We find that fine-tuning Chronos-2 on synthetic utility-maximizing panels substantially improves demand forecasting in a three-good experimental panel, even when most consumers in that panel violate the rationality conditions used to generate the training data. Fine-tuning on GARP-consistent synthetic data reduces bundle prediction error by 17-18\% for horizons $H=5,10,$ and $15$ and by 31\% for horizon $H=1$ relative to zero-shot Chronos-2. 


Together, these results suggest a practical role for economic theory in foundation-model forecasting. Theory provides not only a vocabulary for testing rationality, but also a principled procedure for generating coherent synthetic training data. Models trained on such data inherit the qualitative structure of utility-maximizing behavior as a prior, and that prior proves useful even in populations where rational choice is the exception rather than the rule.

\bibliographystyle{apalike}
\bibliography{references}

\end{document}